\documentclass[sigconf]{acmart}
\usepackage{multirow}
\usepackage{amsmath,amsfonts}
\usepackage{bm}
\usepackage{appendix}

\newtheorem{corollary}{Theorem}

\usepackage{tikz}
\usepackage{pgfplots}
\usepackage{graphicx}
\usepackage{subcaption}
\usepackage{caption}

\definecolor{fig_6_color_1}{RGB}{200,36,35}
\definecolor{fig_6_color_2}{RGB}{248,172,140}
\definecolor{fig_6_color_3}{RGB}{154,201,219}
\definecolor{fig_6_color_4}{RGB}{40,120,181}

\definecolor{fig_7_color_1}{RGB}{199,109,162}
\definecolor{fig_7_color_2}{RGB}{137,131,191}
\definecolor{fig_7_color_3}{RGB}{5,185,226}
\definecolor{fig_7_color_4}{RGB}{50,184,151}

\definecolor{fig_8_color_1}{RGB}{73,108,136}
\definecolor{fig_8_color_2}{RGB}{254,178,180}
\AtBeginDocument{%
  \providecommand\BibTeX{{%
    \normalfont B\kern-0.5em{\scshape i\kern-0.25em b}\kern-0.8em\TeX}}}

\copyrightyear{2024}
\acmYear{2024}
\setcopyright{rightsretained}
\acmConference[SIGIR '24]{Proceedings of the 47th International ACM SIGIR Conference on Research and Development in Information Retrieval}{July 14--18, 2024}{Washington, DC, USA}
\acmBooktitle{Proceedings of the 47th International ACM SIGIR Conference on Research and Development in Information Retrieval (SIGIR '24), July 14--18, 2024, Washington, DC, USA}
\acmDOI{10.1145/3626772.3657742}
\acmISBN{979-8-4007-0431-4/24/07}

\begin{document}

\title{Collaborative Filtering Based on Diffusion Models: \\
Unveiling the Potential of High-Order Connectivity}

\author{Yu Hou}
\orcid{1234-5678-9012}
\affiliation{%
  \institution{Yonsei University}
  \city{Seoul}
  \country{Republic of Korea}
}
\email{houyu@yonsei.ac.kr}

\author{Jin-Duk Park}
\affiliation{%
  \institution{Yonsei University}
  \city{Seoul}
  \country{Republic of Korea}}
\email{jindeok6@yonsei.ac.kr}

\author{Won-Yong Shin}
\affiliation{%
  \institution{Yonsei University}
  \city{Seoul}
  \country{Republic of Korea}
}
\email{wy.shin@yonsei.ac.kr}
\authornote{Corresponding author}
\renewcommand{\shortauthors}{Yu Hou, Jin-Duk Park, \& Won-Yong Shin}

\begin{abstract}

    A recent study has shown that diffusion models are well-suited for modeling the generative process of user--item interactions in recommender systems due to their denoising nature. However, existing diffusion model-based recommender systems do not explicitly leverage high-order connectivities that contain crucial collaborative signals for accurate recommendations. Addressing this gap, we propose \textsf{CF-Diff}, a new diffusion model-based collaborative filtering (CF) method, which is capable of making full use of collaborative signals along with {\it multi-hop} neighbors. Specifically, the forward-diffusion process adds random noise to user--item interactions, while the reverse-denoising process accommodates our own learning model, named cross-attention-guided multi-hop autoencoder (\textsf{CAM-AE}), to gradually recover the original user--item interactions. \textsf{CAM-AE} consists of two core modules: 1) the attention-aided AE module, responsible for precisely learning latent representations of user--item interactions while preserving the model's complexity at manageable levels, and 2) the {\it multi-hop cross-attention} module, which judiciously harnesses high-order connectivity information to capture enhanced collaborative signals. Through comprehensive experiments on three real-world datasets, we demonstrate that \textsf{CF-Diff} is {\bf (a) Superior:} outperforming benchmark recommendation methods, achieving remarkable gains up to 7.29\% compared to the best competitor, {\bf (b) Theoretically-validated:} reducing computations while ensuring that the embeddings generated by our model closely approximate those from the original cross-attention, and {\bf (c) Scalable:} proving the computational efficiency that scales {\it linearly} with the number of users or items.

\end{abstract}





\maketitle

\section{Introduction}

Diffusion models \cite{sohl2015deep, ho2020denoising} have become one of recent emerging topics thanks to their state-of-the-art performance in various domains, including computer vision \cite{ho2020denoising,rombach2022high,dhariwal2021diffusion}, natural language processing \cite{austin2021structured,savinov2021step}, and multi-modal deep learning \cite{avrahami2022blended, saharia2022photorealistic}. Diffusion models, categorized as deep generative models, gradually perturb the input data by adding random noise in the forward-diffusion process and then recover the original input data by learning in the reverse-denoising process, step by step. Due to their denoising nature, diffusion models align well with recommender systems, which can be viewed as a denoising process because user--item historical interactions are naturally noisy and diffusion models can learn to recover the original interactions based on corrupted ones \cite{ wu2016collaborative,li2017collaborative,wang2023diffusion}. Recent efforts have verified the effectiveness of diffusion models for sequential recommendations \cite{yang2023generate, liu2023diffusion, wu2023diff4rec, li2023diffurec}, where the process of modeling sequential item recommendations mirrors the step-wise process of diffusion models. However, the application of diffusion models to recommender systems has yet been largely underexplored. 



\begin{figure}
    \centering
    
    \begin{subfigure}[b]{0.4\linewidth} 
        \centering
        \includegraphics[width=\linewidth]{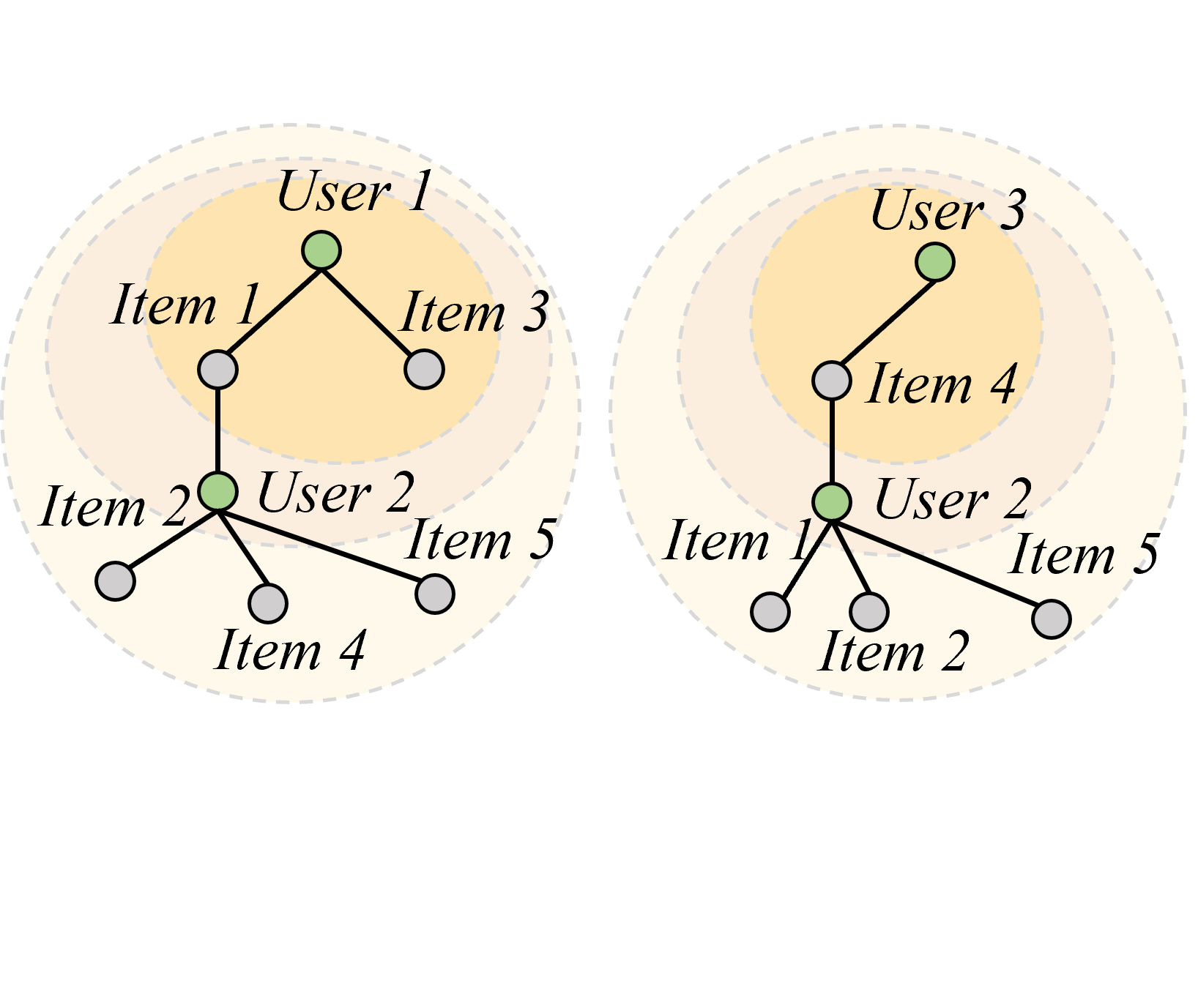} 
        \captionsetup{font={small,stretch=0.5}, skip=2pt,textfont=normalfont,labelfont=normalfont}
        \caption{Multi-hop neighbors}
        \label{fig:sub1}
    \end{subfigure}
    \hspace{0.1cm}
    \begin{subfigure}[b]{0.57\linewidth} 
        \centering
        \includegraphics[width=\linewidth]{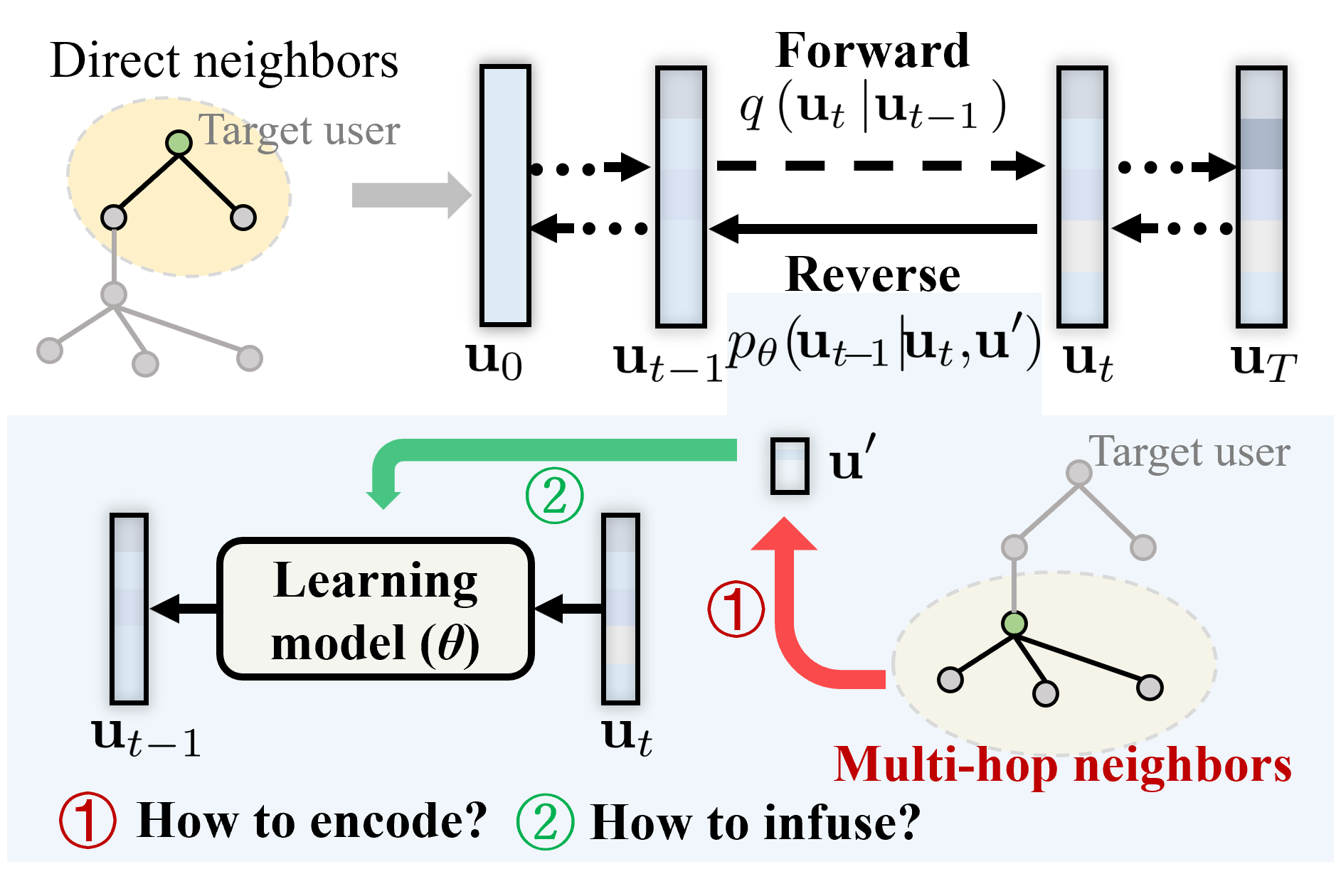} 
        \captionsetup{font={small,stretch=0.5}, skip=2pt,textfont=normalfont,labelfont=normalfont}
        \caption{New challenges}
        \label{fig:sub2}
    \end{subfigure}
    \captionsetup{skip=5pt}
    \caption{Illustration showing (a) neighbors of {\em User 1} and {\em User 3} up to 3 hops and (b) how such high-order connectivity information can be potentially encoded and infused into the diffusion model-based learning system. Here, $\left\{ {{\bf{u}}_0 , \cdots ,{\bf{u}}_T } \right\} $ are the encoded information of direct user--item interactions at each step, and ${\bf u}'$ is the encoded high-order connectivity information.}
    \label{fig:motivations}
\vspace{-2.5em}
\end{figure}

On one hand, one of the dominant techniques used in recommender systems is collaborative filtering (CF), where attention has been paid to model-based approaches including matrix factorization (MF) \cite{koren2009matrix, yang2018hop} and deep learning \cite{he2017neural, liang2018variational, zhou2018deep,wang2019neural,he2020lightgcn,ParkL0S23} ({\it e.g.}, graph neural networks (GNNs) \cite{wang2019neural,he2020lightgcn,ParkL0S23}). CF-based recommender systems have achieved great success in many real-world applications, due to their simplicity, efficiency, and effectiveness, while aiming to learn multi-hop relationships among users and items. For example, the message passing mechanism in GNNs, being increasingly used in the tasks of recommendation, captures collaborative signals in high-order connectivities by aggregating features of neighbors. Figure \ref{fig:motivations}a illustrates the multi-hop neighbors used for CF with an example involving two users. It is seen that, although {\it User 1} and {\it User 3} have different direct interactions, they share similar 2-hop ({\it User 2}) and 3-hop ({\it Item 2}, {\it Item 5}) neighbors, which implies that {\it User 1} ({\it resp.} {\it User 3}) is highly likely to prefer {\it Item 4} consumed by {\it User 3} ({\it resp.} {\it Item 1} and {\it Item 3} consumed by {\it User 1}). 

On the other hand, unlike the existing CF techniques using MF and GNNs, it is not straightforward to grasp how to exploit such high-order connectivity information from a {\it diffusion model}'s perspective, as shown in Figure \ref{fig:motivations}b. Recent studies on diffusion model-

\noindent
based recommender systems \cite{walker2022recommendation, wang2023diffusion, yang2023generate, liu2023diffusion, wu2023diff4rec, li2023diffurec} often overlooked the exploration of multi-hop similarity/proximity among nodes, albeit the core mechanism of CF in achieving satisfactory performance. In this context, even with recent attempts to develop recommender systems via diffusion models \cite{walker2022recommendation, wang2023diffusion, yang2023generate, liu2023diffusion, wu2023diff4rec, li2023diffurec}, a natural question arising is: ``how can {\it high-order connectivity} information be efficiently and effectively incorporated into recommender systems based on diffusion models?''. To answer this question, we would like to outline the following two {\bf design challenges}:
\begin{itemize}
    \item {\bf C1.} how to ensure the complexity of the learning model (to be designed) at an acceptable level even when including high-order connectivity information;
    \item {\bf C2.} how to judiciously link the high-order connectivity information with the direct user--item interactions under a diffusion-model framework.
\end{itemize}

It is worth noting that leveraging direct user--item interactions ({\it i.e.}, direct neighbors) of each individual is rather straightforward so that diffusion models can learn the distribution of these interactions (see, {\it e.g.},  \cite{wang2023diffusion} for such an attempt). However, the exploration of high-order collaborative signals among users and items inevitably poses technical challenges. First, the infusion of high-order connectivity information may lead to an increased memory and computational burden, as training diffusion models is known to be quite expensive in terms of space and time \cite{sohl2015deep, ho2020denoising}. This complexity issue will be severe with an increasing number of users and items. Second, injecting high-order connectivities in an explicit manner into a learning system within a diffusion-model framework is technically abstruse. As shown in Figure 1b, while direct user--item interactions can be readily fed to the diffusion model-based learning system, the accommodation of high-order collaborative signals necessitates a complex and challenging integration task.

To address these aforementioned challenges, we make the first attempt towards developing a {\it lightweight} \underline{CF} method based on \underline{diff}usion models, named {\bf \textsf{CF-Diff}}. 

{\bf (\underline{Idea 1})} The proposed \textsf{CF-Diff} method naturally involves two distinct processes, the forward-diffusion process and the reverse-denoising process. The forward-diffusion process gradually adds random noise to the individual user--item interactions, while the reverse-denoising process aims to gradually recover these interactions by infusing {\it high-order connectivities}, achieved through our proposed learning model to be specified later.

{\bf (\underline{Idea 2})} As one of our main contributions, we next design an efficient yet effective learning model for the reverse-denoising process, dubbed \underline{c}ross-\underline{a}ttention-guided \underline{m}ulti-hop \underline{a}uto\underline{e}ncoder (\textsf{CAM-AE}), which is capable of infusing and learning high-order connectivities without incurring additional computational costs and scalability issues. Our \textsf{CAM-AE} model consist of three primary parts: a high-order connectivity encoder, an {\it attention-aided AE} module, and a {\it multi-hop cross-attention} module. First, we initially pre-process the user--item interactions in the sense of extracting and encoding `per-user' connectivity information from pre-defined multi-hop neighboring nodes. Next, we incorporate the attention-aided AE module into \textsf{CAM-AE} to precisely learn latent representations of the noisy user--item interactions while preserving the model's complexity at manageable levels by controlling the dimension of latent representations (\underline{solving the challenge {\bf C1}}). Lastly, inspired by conditional diffusion models \cite{rombach2022high}, we incorporate the multi-hop cross-attention module into \textsf{CAM-AE} since high-order connectivity information can be seen as a {\it condition} for denoising the original user--item interactions. This module takes advantages of the conditional nature of these connectivities while connecting with the direct user--item interactions in the reverse-denoising process, thereby enriching the collaborative signal (\underline{solving the challenge {\bf C2}}).

Our main contributions are summarized as follows:
\begin{itemize} 
    \item {\bf Novel methodology}: We propose \textsf{CF-Diff}, a novel diffusion model-based CF method featuring our specially designed learning model, \textsf{CAM-AE}. This model is composed of 1) the encoder of high-order connectivity information, 2) the attention-aided AE module primarily designed for preserving the model’s complexity at manageable levels, and 3) the multi-hop cross-attention module for accommodating high-order connectivity information. 
    \item {\bf Extensive evaluations}: Through comprehensive experimental evaluations on three real-world benchmark datasets, including two large-scale datasets, we demonstrate (a) the superiority of \textsf{CF-Diff}, showing substantial gains up to 7.29\% in terms of NDCG@10 compared to the best competitor, (b) the effectiveness of core components in \textsf{CAM-AE}, and (c) the impact of multi-hop neighbors in \textsf{CF-Diff}.
    \item {\bf Theoretical findings}: We theoretically prove that (a) our learning model's embeddings closely approximate those from the (computationally more expensive) original cross-attention, and (b) the model's computational complexity scales {\it linearly} with the maximum between the number of users and the number of items. This is further supported by empirical verifications, confirming the scalability of \textsf{CF-Diff}.
\end{itemize}

\section{Methodology}

\subsection{Notations}

Let $u\in\mathcal{U}$ and $i\in\mathcal{I}$ denote a user and an item, respectively, where $\mathcal{U}$ and $\mathcal{I}$ denote the sets of all users and all items, respectively. Historical interactions of a user $u \in \mathcal{U}$ with items are represented as a binary vector ${\bf u} \in \left\{ {0,1} \right\}^{\left| \mathcal{I} \right|} $ whose $i$-th entry is 1 if there exists implicit feedback (such as a click or a view) between user $u$ and item $i \in \mathcal{I}$, and 0 otherwise.\footnote{The unbolded $u$ represents a user, while the bolded ${\bf u}$ represents a certain  user's interaction vector as utilized in the proposed method.}
    
    

\subsection{Overview of \textsf{CF-Diff}}
We describe the methodology of \textsf{CF-Diff}, a new diffusion model-based CF method that is capable of reflecting high-order connectivity information, revealing co-preference patterns between users and items, for accurate recommendations. We recall that recent recommendation methods using diffusion models \cite{walker2022recommendation, wang2023diffusion, yang2023generate, liu2023diffusion, wu2023diff4rec, li2023diffurec} focus primarily on leveraging only the direct user--item interactions and overlook the collaborative signal in high-order connectivities during training. Our study aims to fill this gap by infusing high-order connectivity information into the proposed method, which poses two main design challenges that we mentioned earlier: preserving the learning model's complexity at an acceptable level ({\bf C1}) and learning complex high-order connectivities at a fine-grained level ({\bf C2}).

\begin{figure}
    \centering
    \includegraphics[width=1.0\linewidth]{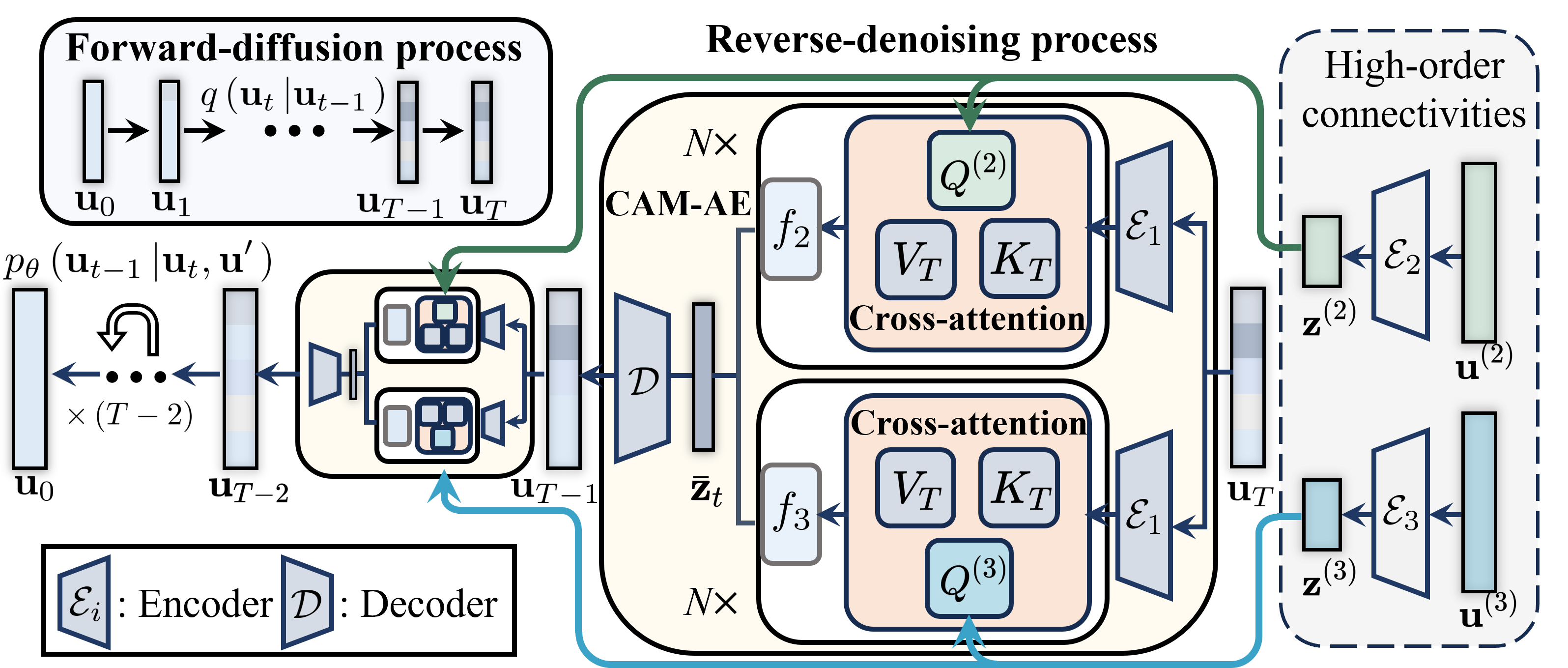}
    \captionsetup{skip=5pt}
    \caption{The schematic overview of \textsf{CF-Diff} when both $2$-hop and $3$-hop neighboring nodes are taken into account.}
    \label{fig:overview} 
\vspace{-1em}
\end{figure}

To tackle these challenges, as a core module of \textsf{CF-Diff}, we develop an innovative learning model, \textsf{CAM-AE}. In the \textsf{CAM-AE} model, we propose to use a {\it multi-hop cross-attention} mechanism to infuse multi-hop neighborhood information from the target user during training, thereby enriching the collaborative signal, which however causes additional computational costs. To counter this, we next employ an {\it attention-aided AE} module, enabling to preserving the model's complexity at manageable levels.

Note that diffusion models can be viewed as partitioning the denoising process of an AE into a series of finer sub-processes \cite{deja2022analyzing,huang2021variational}, which can capture more delicate recovery details. Since \textsf{CF-Diff} is built upon such diffusion models, it naturally involves two distinct processes, namely the forward-diffusion process and the reverse-denoising process, achieved with a tailored neural network architecture in \textsf{CAM-AE}. The schematic overview of the \textsf{CF-Diff} method is illustrated in Figure \ref{fig:overview}, and each process in \textsf{CF-Diff} is summarized as follows.

\begin{enumerate}
    \item {\bf Forward-diffusion process (Section 2.3)}: The forward diffusion, aligning with standard diffusion models, gradually adds Gaussian noise to the user--item historical interactions, as shown in the upper left part of Figure \ref{fig:overview}.
    \item {\bf Reverse-denoising process (Section 2.4)}: We aim to gradually recover the original user--item interactions from noisy ones. This is achieved by using the proposed learning model, \textsf{CAM-AE} (to be specified Section 3), which infuses high-order connectivities to iteratively guide the reverse-denoising process. To bridge the historical one-hop interactions and multi-hop neighbors, our \textsf{CAM-AE} model integrates an attention-aided AE with a cross-attention architecture, progressively recovering user--item interactions by leveraging high-order connectivity information (see the right part of Figure \ref{fig:overview}).
\end{enumerate}

\subsection{Forward-Diffusion Process}

We denote the initial state of a specific user $u \in \mathcal{U}$ as ${\bf{u}}_0  = {\bf u}$.\footnote{For notational convenience, since each user $u$ experiences the forward-diffusion and reverse-denoising processes independently, we do not use the user index in $u$ unless it causes any confusion.} In the forward-diffusion process, we gradually insert Gaussian noise in the initial user--item interactions ${\bf u}_0$ over $T$ steps, producing a sequence of noisy samples ${\bf u}_1 , \ldots ,{\bf u}_T $, denoted as ${\bf u}_{1:T} $ (see Figure \ref{fig:overview}), which can be modeled as 
\begin{equation}
    q\left( {{\bf{u}}_{1:T} \left| {{\bf{u}}_0 } \right.} \right) = \prod\limits_{t = 1}^T q\left( {{\bf{u}}_t \left| {{\bf{u}}_{t - 1} } \right.} \right),
\end{equation}
where
\begin{equation}
 q\left( {{\bf u}_t \left| {{\bf u}_{t - 1} } \right.} \right) = {\mathcal{N}}\left( {{\bf u}_t ;\sqrt {1 - \beta _t } {\bf u}_{t - 1} ,\beta _t {\bf I}} \right) 
\end{equation}
represents the transition of adding noise from states ${\bf u}_{t-1}$ to ${\bf u}_t$ via a Gaussian distribution \cite{sohl2015deep, ho2020denoising}. Here, $t \in \left\{ {1, \ldots ,T} \right\} $ refers to the diffusion step; $\mathcal{N}$ denotes the Gaussian distribution; and $\beta _t  \in \left( {0,1} \right) $ controls the Gaussian noise scales added at each time step $t$. To generate the noisy sample ${\bf u}_t$ from $q\left( {{\bf u}_t \left| {{\bf u}_{t - 1} } \right.} \right) $, we employ the reparameterization trick \cite{kingma2013auto}, expressed as ${\bf u}_t  = \sqrt {1 - \beta _t } {\bf u}_{t - 1}  + \sqrt {\beta _t } \varepsilon _{t - 1} $, where $\varepsilon _{t - 1}  \sim \mathcal{N}\left( {\bf{0},\bf{I}} \right) $. This process is iteratively applied until we obtain the final sample ${\bf u}_T$ at time step $T$.

It is noteworthy that, in contrast to existing diffusion models, our approach focuses on adding noise to user--item interactions from a {\it single user}'s perspective, which originates from the nature of the denoising process in variational AE (VAE)-based CF \cite{li2017collaborative}.

\subsection{Reverse-Denoising Process}

In the reverse-denoising process, the estimation of the distribution $q\left( {{\bf u}_{t - 1} \left| {{\bf u}_t } \right.} \right) $ is technically not easy as it requires using the entire dataset. Therefore, a neural network model $p_\theta $ is employed to approximate such conditional probabilities \cite{ho2020denoising}. Starting from ${\bf u}_T$, the reverse-denoising process gradually recovers ${\bf u}_{t-1}$ from ${\bf u}_t$ via the denoising transition step. However, only relying on user--item interactions do not ensure the high-quality recovery for CF-based recommendations, as high-order connectivity information plays an important role in guaranteeing state-of-the-art performance of CF, as shown in Figure \ref{fig:motivations}a.

To address this, we integrate multi-hop neighbors of the target user $u$ (denoted as ${\bf u}'$) into our learning model, thereby enhancing recommendation accuracies. This differs from original diffusion models, which focus on denoising solely from noisy samples ({\it i.e.}, $p_\theta  \left( {{\bf u}_{t - 1} \left| {{\bf u}_t } \right.} \right)$ in \cite{ho2020denoising}). In other words, our approach not only denoises from noisy samples but also enriches the denoising process by exploiting high-order connectivities. The denoising transition via the Gaussian distribution is formulated as follows \cite{ho2020denoising, rombach2022high}:
\begin{equation}
    p_\theta  \left( {{\bf{u}}_{0:T} } \right) = p\left( {{\bf{u}}_T } \right)\prod\limits_{t = 1}^T p_\theta  \left( {{\bf{u}}_{t - 1} \left| {{\bf{u}}_t ,{\bf{u}}^\prime  } \right.} \right),
\end{equation}
where
\begin{equation}
 p_\theta  \left( {{\bf u}_{t - 1} \left| {{\bf u}_t, {\bf u}' } \right.} \right) = \mathcal{N}\left( {{\bf u}_{t - 1} ;\bm{\mu} _\theta  \left( {{\bf u}_t , {\bf u}', t} \right),\bm{\Sigma} _\theta  {\left( {{\bf u}_t , {\bf u}', t} \right)} } \right). 
\label{eq:reverse}
\end{equation}
Here, ${\bm{\mu} _\theta  \left( {{\bf u}_t , {\bf u}', t} \right)} $ and ${\bm{\Sigma}_\theta  {\left( {{\bf u}_t, {\bf u}', t} \right)} } $ are the mean and covariance of the Gaussian distribution predicted by the neural network with learnable parameters $\theta $. Besides, to maintain training stability and simplify calculations, we ignore learning of ${\bm{\Sigma}_\theta  {\left( {{\bf u}_t , {\bf u}', t} \right)} } $ in Eq. (\ref{eq:reverse}) and set $\bm{\Sigma}_\theta  {\left( {{\bf u}_t ,{\bf u}',t} \right)}  = \beta _t {\bf I} $ by following \cite{ho2020denoising}. After leaning the mean ${\bm{\mu} _\theta  \left( {{\bf u}_t , {\bf u}', t} \right)} $ in the model, we can obtain the recovered ${\bf u}_{t-1}$ by sampling from $p_\theta  \left( {{\bf u}_{t - 1} \left| {{\bf u}_t, {\bf u}' } \right.} \right)$. This process is iteratively applied until we obtain an estimate of the original sample ${\bf u}_0$.

The neural network architecture of \textsf{CAM-AE} is designed in the sense of judiciously infusing high-order connectivities in the reverse-denoising process. To this end, \textsf{CAM-AE} consists of two key components: 1) an {\it attention-aided AE} module precisely learns latent representations of the noisy user--item interactions, helping preserve the complexity manageable (solving the challenge {\bf C1}), and 2) a {\it multi-hop cross-attention} module, which accommodates high-order connectivity information to facilitate the reverse-denoising process, thus capturing the enriched collaborative signal (solving the challenge {\bf C2}).

\section{Learning Model: \textsf{CAM-AE}}

In this section, we elaborate on the proposed \textsf{CAM-AE} model, comprising an attention-aided AE module and a multi-hop cross-attention module. After showing how to extract and encode multi-hop neighborhood information for a given bipartite graph, we describe implementation details of each module in \textsf{CAM-AE}. We then explain how to optimize our learning model. Finally, we provide analytical findings, which theoretically validate the efficiency of \textsf{CAM-AE}. 

\subsection{High-Order Connectivity Encoder}

To extract multi-hop neighbors of a given user, we may use a bipartite graph constructed by establishing edges based on all user--item interactions. However, using such a bipartite graph will result in a huge memory and computational burden during training. To solve this practical issue, we pre-process the user--item interactions in such a way of initially extracting multi-hop neighbors of a user. This extracted `{\it per-user}' connectivity information is then made available in the reverse-denoising process to assist recovery of the original user--item interactions  (see Figure \ref{fig:overview}).

\begin{figure}
    \centering
    \includegraphics[width=0.9\linewidth]{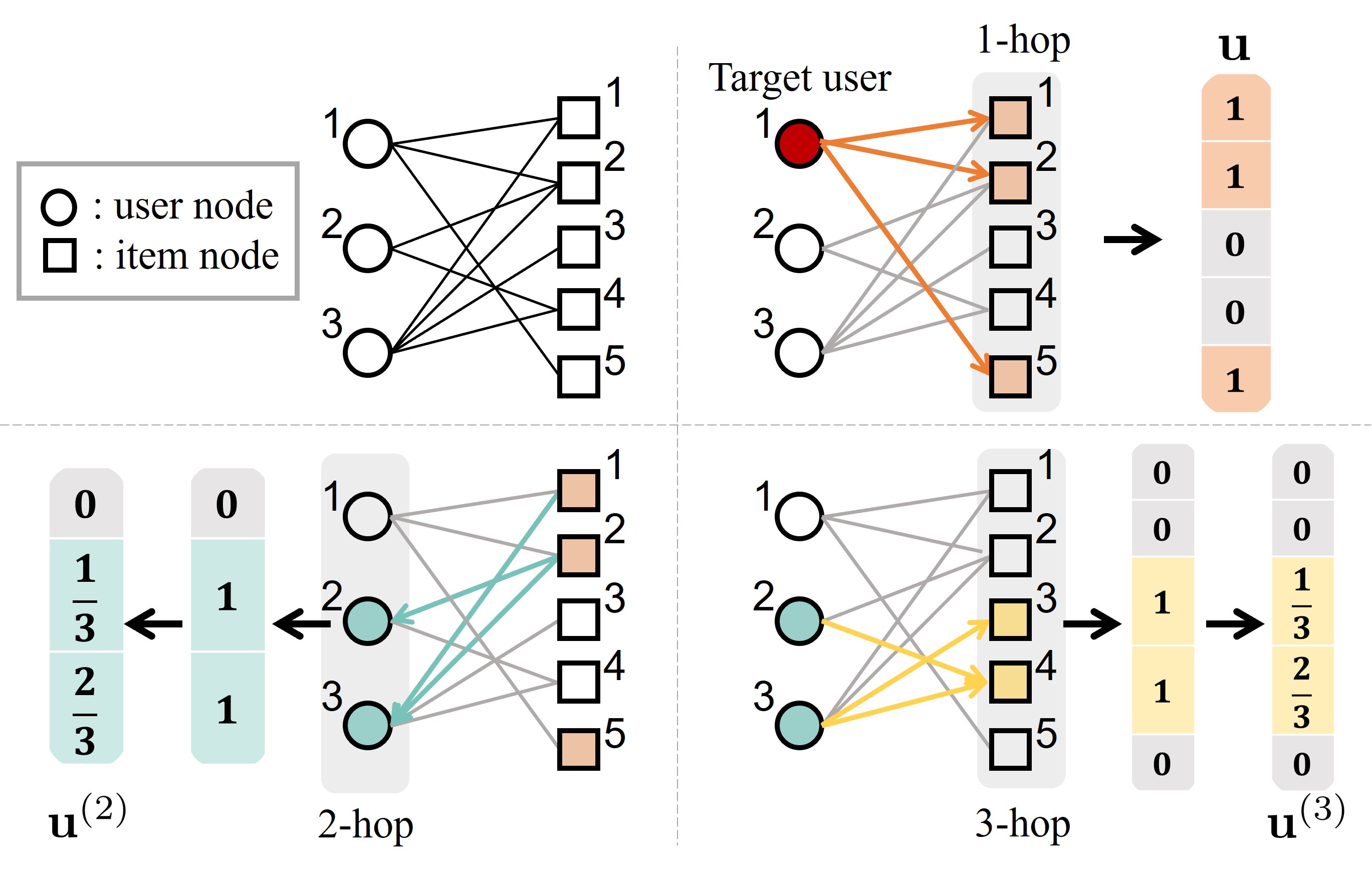}
    \captionsetup{skip=-2pt} 
    \caption{Extraction and encoding of 2-hop and 3-hop neighbors of the target user ({\em User 1}) as well as direct neighbors for a given bipartite graph.}
    \label{fig:high_order} 
\vspace{-1em}
\end{figure}

Given a target user’s historical interactions {\bf u}, we explain how to explore multi-hop neighbors along paths within the user--item bipartite graph. In our study, we encode high-order connectivity information ({\it i.e.}, high-order collaborative signals) up to $H$-hop neighbors as in the following form: 
\begin{equation}
    {\bf{u}}^\prime   = \left[ {{\bf{u}}^{\left( 2 \right)} , \ldots ,{\bf{u}}^{\left( H \right)} } \right], 
\label{eq:high-order}
\end{equation}
where 
\begin{equation}
    {\bf u}^{\left( h \right)}  = \frac{1}{N_{h-1,h}} {\bf r}\left( {\mathcal{G}\left( {u,h} \right),{\bf c}^{(h)} } \right)
\end{equation}
for $h=2,\cdots,H$.\footnote{If $h$ is even, then $\left| {{\bf{u}}^{\left( h \right)} } \right| = \left| \mathcal{U} \right|$. Otherwise, $\left| {{\bf{u}}^{\left( h \right)} } \right| = \left| \mathcal{I} \right|$. However, to tractably handle ${\bf u}'$, we can set the dimensionality of each ${\bf u}^{(h)}$ to $\max \left\{ {\left| \mathcal{U} \right|,\left| \mathcal{I} \right|} \right\}$.} Here, ${\bf r}(\cdot,\cdot)$ is the vector-valued function returning a multi-hot encoded vector where one is assigned only to the elements corresponding to $h$-hop neighbors of user $u$; $\mathcal{G}\left( {u,h} \right) $ indicates the set of $h$-hop neighbors of user $u$; ${\bf c}^{(h)}\in \mathbb{R}^{|\mathcal{G}(u,h)| \times 1}$ is the integer vector, each of which represents the number of incoming links from $(h-1)$-hop neighbors of user $u$ to each of $h$-hop neighbors; and $N_{h-1,h}$ is the total number of interactions between $(h-1)$-hop and $h$-hop neighbors of of user $u$. Now, let us show an explicit form of encoded $h$-hop neighborhood information ${\bf u}^{(h)}$ along with the following example. 
\begin{example}
Consider the target user ({\it User 1}) in the user--item bipartite graph consisting of 3 users and 5 items, as illustrated in Figure \ref{fig:high_order}. Here, it follows that ${\bf u} = \left[ {1\;1\;0\;0\;1} \right]^T $ as {\it User 1} has interacted with {\it Item 1}, {\it Item 2}, and {\it Item 5}. Since the 2-hop neighbors of {\it User 1} are {\it User 2}, {\it User 3} and {\it User 3} has two incoming links, we have ${\bf{u}}^{(2)}  = [0\;\frac{1}{3}\;\frac{2}{3}]^T $ normalized to the total number of interactions at the second hop. Similarly, we obtain ${\bf{u}}^{(3)}  = [0\;0\;\frac{1}{3}\;\frac{2}{3}\;0]^T $.
\end{example}

\subsection{Attention-Aided AE Module}

VAE-based CF \cite{li2017collaborative} shows great potential in capturing underlying patterns by encoding user--item interactions into a latent space. Similarly, in the \textsf{CAM-AE} model, we would like to design lightweight encoders to project the user--item interactions into a latent space, aiming to capture high-level patterns while keeping the computations manageable by controlling the latent dimension. This design principle enables us to solve the challenge {\bf C1}. 

In \textsf{CAM-AE}, the attention-added AE module involves {\it hop-specific} encoders. As illustrated in Figure \ref{fig:overview}, an encoder $\mathcal{E}_1 \left(  \cdot  \right) $ is adopted to project user $u$'s noisy interactions ${\bf u}_t$ into a latent space, represented by the latent embedding ${\bf z}_t  \in \mathbb{R}^{^{k \times 1} } $ with its dimensionality $k$. Likewise, another hop-specific encoder $\mathcal{E}_h \left(  \cdot  \right)$ generates embeddings for the encoded information of $h$-hop neighbors of user $u$, ${{\bf{u}}^{\left( h \right)} } $, yielding  ${\bf z}^{\left( h \right)}  \in \mathbb{R}^{^{k \times 1} } $. Similarly as in \cite{wang2020linformer}, these two encoders $\mathcal{E}_1 \left(  \cdot  \right)$ and $\mathcal{E}_h \left(  \cdot  \right)$ are implemented as {\it linear} transformations, which are formally expressed as
\begin{equation}
   {{\bf{z}}_t  = {\mathcal{E}}_1 \left( {{\bf{u}}_t } \right) = {\bf E}_1  {{\bf{u}}_t } , } 
\label{eq:encoder1}
\end{equation}
\begin{equation}
   {{\bf{z}}^{\left( h \right)}  = {\mathcal{E}}_h \left( {{\bf{u}}^{\left( h \right)} } \right) = {\bf E}_h {{\bf{u}}^{\left( h \right)} }  ,} 
\label{eq:encoder2}
\end{equation}
where ${\bf E}_1  \in \mathbb{R}^{^{k \times {\left| \mathcal{I} \right|}} } $ and ${\bf E}_h  \in \mathbb{R}^{^{k \times {\left| {\bf{u}}^{\left( h \right)} \right|}} } $ represents the transformation matrices. Figure \ref{fig:overview} illustrates the case where the embeddings ${\bf z}^{(2)}$ and ${\bf z}^{(3)}$ of both 2-hop and 3-hop neighbors of a target user are generated.\footnote{Although the example in Figure \ref{fig:overview} deals with up to 3-hop neighbors, it is straightforward to extend our module to the case of leveraging general $h$-hop neighbors.} We can preserve the \textsf{CAM-AE} model's complexity at manageable levels through these linear transformations that reduce the dimension of latent representations.

We turn to addressing a decoder $\mathcal{D} \left(  \cdot  \right)$, which is adopted to recover the mean value of $p_\theta  \left( {{\bf u}_{t - 1} \left| {{\bf u}_t, {\bf u}' } \right.} \right) $ using the embedding, denoted as $\bar{\bf z}_t$, as input that is returned by the multi-hop cross-attention module (to be specified in Section 3.3), as depicted in Figure \ref{fig:overview}. The decoder is formulated as follows: 
\begin{equation}
    \hat {\bm \mu} _\theta   = \mathcal{D}\left( {\bar {\bf z}_t } \right) = {\bf D} \bar {\bf z}_t, 
\label{eq:decoder}
\end{equation}
where ${\bf D}\in \mathbb{R}^{^{{\left| \mathcal{I} \right|} \times k} }$ is the transformation matrix in the decoder. Then, ${\bf u}_{t-1}$ can be sampled from $p_\theta  \left( {{\bf u}_{t - 1} \left| {{\bf u}_t, {\bf u}' } \right.} \right) = \mathcal{N}\left( {{\bf u}_{t - 1} ; \hat {\bm \mu} _\theta , \beta _t {\bf I} } \right)$.

\subsection{Multi-Hop Cross-Attention Module}
The \textsf{CAM-AE} model is enlightened by conditional diffusion models \cite{rombach2022high}, which achieved impressive success in various fields by using the cross-attention mechanism \cite{vaswani2017attention} to integrate additional conditions. In \textsf{CAM-AE}, high-order connectivity information in Eq. (\ref{eq:high-order}) can be regarded as a {\it condition} for denoising the original user--item interactions ${\bf u}_0$, following the principle of conditional diffusion models  \cite{rombach2022high}. In this study, to effectively infuse high-order connectivities into our learning model, we propose the multi-hop cross-attention module. This module judiciously harnesses the conditional nature of these connectivities while connecting with the direct user--item interactions in the reverse-denoising process. This design principle is established to fundamentally solve the challenge {\bf C2}. 

In the multi-hop cross-attention module, we start by expanding the dimension of ${\bf z}_t  \in \mathbb{R}^{^{k \times 1} } $ and ${\bf z}^{\left( h \right)}  \in \mathbb{R}^{^{k \times 1} } $ ({\it i.e.}, the output embeddings of encoders $\mathcal{E}_1$ and $\mathcal{E}_h$) to obtain ${\bf v}_t  \in \mathbb{R}^{^{k \times d} } $ and ${\bf q}^{\left( h \right)}  \in \mathbb{R}^{^{k \times d} } $ for improving the {\it expressiveness}. This expansion can be implemented as ${\bf v}_t  = {\bf{z}}_t {\bf E}^v $ and ${\bf q}^{\left( h \right)}  = {\bf{z}}^{\left( h \right)} {\bf E}^q $, where ${\bf E}^v  \in \mathbb{R}^{1 \times d} $ and ${\bf E}^q  \in \mathbb{R}^{1 \times d} $ are the transformation matrices with $d$ being the expended dimensionality. Then, the resulting embedding of $h$-hop neighbors of user $u$, ${\bf q}^{\left( h \right)}$, is integrated into ${\bf v}_t$ using the multi-hop cross-attention module:
\begin{equation}
    Attention_h\left( {{\bf Q}^{\left( h \right)},{\bf K}_t,{\bf V}_t} \right) := softmax \left( {\frac{{{\bf Q}^{\left( h \right)}{\bf K}_t^T}}{{\sqrt d }}} \right){\bf V}_t,
    \label{eq:attention}
\end{equation}
where ${\bf Q}^{\left( h \right)}\! = \!{\bf q}^{\left( h \right)} {\bf W}_\theta ^Q $, ${\bf K}_t \! =\! {\bf v}_t {\bf W}_\theta ^K $, and ${\bf V}_t \!=\! {\bf v}_t {\bf W}_\theta ^V $; and $softmax \left(  \cdot  \right) $ is the softmax function. Here, $\left\{ {{\bf W}_\theta ^Q ,{\bf W}_\theta ^K ,{\bf W}_\theta ^V } \right\} \in \mathbb{R}^{^{d \times d} } $ are trainable parameters. Figure \ref{fig:overview} includes the multi-hop cross-attention module (see the light red blocks in the reverse-denoising process) when 2-hop and 3-hop neighbors of the target user are taken into account.

Due to the fact that the aforementioned process is basically built upon linear transformations that lack the ability to capture the intrinsic data complexity, a per-hop forward operation $f_h\left(  \cdot  \right) $ using {\it non-linear} transformations is applied to $Attention_h$ \cite{vaswani2017attention}. We stack $N$ identical layers, each consisting of cross-attention and non-linear transformation, with the output from the last layer aggregated to form ${\bar {\bf{z}}}_t  \in \mathbb{R}^{k \times d} $, calculated as
\begin{equation}
    {\bar {\bf z}}_t  = \sum\limits_{h = 2}^H {\alpha _h f_h \left( {Attention_h } \right)},
\label{eq:output}
\end{equation}
where $\sum\limits_{h = 2}^H {\alpha _h  = 1}$; $\alpha _h$ is the weight balancing among different $f_h(\cdot)$'s specific to hop $h$; and $H$ is the number of hops. Finally, $\bar {\bf{z}}_t $ is the input of the decoder $\mathcal{D} \left(  \cdot  \right) $ in Eq. (\ref{eq:decoder}).

It is worthwhile to note that both ${\bf v}_t$ and ${\bf q}^{\left( h \right)} $ originate from the same user, offering two different perspectives of the same data. This dual perspective is beneficial for precisely capturing the collaborative signal. In other words, through the cross-attention mechanism  in \textsf{CAM-AE}, high-order connectivities can significantly improve the reverse-denoising process, thereby ultimately enhancing recommendation accuracies.

\subsection{Optimization}


In our learning model, the denoising transition $p_\theta  \left( {{\bf{u}}_{t - 1} \left| {{\bf{u}}_t ,{\bf{u}}^\prime  } \right.} \right) =  \mathcal{N}\left( {{\bf{u}}_{t - 1} ;} \right.$ $\left. {{\bm \mu} _\theta  \left( {{\bf{u}}_t ,{\bf{u}}^\prime  ,t} \right),\beta _t {\bf{I}}} \right) $ is forced to approximate the tractable distribution $q\left( {{\bf{u}}_{t - 1} \left| {{\bf{u}}_t ,{\bf{u}}_0 } \right.} \right) = \mathcal{N}\left( {{\bf u}_{t - 1} ;\tilde {\bm \mu} \left( {{\bf u}_t ,{\bf u}_0 } \right),\beta _t  {\bf I}} \right) $ (note that the mean $\tilde {\bm \mu} \left( {{\bf u}_t ,{\bf u}_0 } \right) $ can be computed via Bayes' rule as shown in \cite{ho2020denoising}: $ q\left( {{\bf{u}}_{t - 1} \left| {{\bf{u}}_t ,{\bf{u}}_0 } \right.} \right) = q\left( {{\bf{u}}_t \left| {{\bf{u}}_{t - 1} ,{\bf{u}}_0 } \right.} \right)\frac{{q\left( {{\bf{u}}_{t - 1} \left| {{\bf{u}}_0 } \right.} \right)}}{{q\left( {{\bf{u}}_t \left| {{\bf{u}}_0 } \right.} \right)}} $). Following this approximation, we can generate ${\bf u}_{t-1}$ from ${\bf u}_t$ progressively until ${\bf u}_0$ is reconstructed. Figure \ref{fig:overview} visualizes a single denoising step from ${\bf u}_T$ to ${\bf u}_{T-1}$, which is repeated $T$ times to obtain ${\bf u}_0$.

To optimize the parameter $\theta $, our model aims at minimizing the variational lower bound (VLB) \cite{kingma2013auto, ho2020denoising} for the observed user--item interactions ${\bf u}_0$ alongside the following loss:
\begin{equation}
    \mathcal{L}_\text{VLB}  = \mathcal{L}_0  + \sum\nolimits_{t = 2}^T {\mathcal{L}_{t - 1} } ,
\end{equation}
where $\mathcal{L}_0 = \mathbb{E}_q \left[ { - \log p_\theta  \left( {{\bf{u}}_0 \left| {{\bf{u}}_1 ,{\bf{u}}^\prime  } \right.} \right)} \right] $ is the reconstruction term to recover the original interactions ${\bf u}_0$; and $\mathcal{L}_{t-1}$ is the denoising matching term, regulating $p_\theta  \left( {{\bf u}_{t - 1} \left| {{\bf u}_t , {\bf u}'} \right.} \right) $ to align with the tractable distribution $q\left( {{\bf{u}}_{t - 1} \left| {{\bf{u}}_t ,{\bf{u}}_0 } \right.} \right) $, served as the ground truth, and is given by
\begin{equation}
    \begin{array}{l}
    \mathcal{L}_{t-1}  = \mathbb{E}_q \left[ {D_\text{KL} \left( {q\left( {{\bf u}_{t - 1} \left| {{\bf u}_t ,{\bf u}_0 } \right.} \right)\left\| {p_\theta  \left( {{\bf u}_{t - 1} \left| {{\bf u}_t ,{\bf u}'} \right.} \right)} \right.} \right)}\right] \\ 
    \;\;\;\;\;\;\;\;= \mathbb{E}_q \left[ {\frac{1}{{2\beta_t}}\left[ {\left\| {{\bm \mu} _\theta  \left( {{\bf u}_t ,{\bf u}',t} \right) - \tilde {\bm \mu} \left( {{\bf u}_t ,{\bf u}_0} \right)} \right\|^2 } \right]} \right], \\ 
    \end{array}
\end{equation}
where $D_\text{KL}(\cdot \| \cdot)$ denotes the Kullback–Leibler (KL) divergence between two distributions.

\subsection{Theoretical Analyses}

In this subsection, we are interested in theoretically showing the efficiency of the \textsf{CAM-AE} model. In \textsf{CAM-AE}, we use an AE to generate embeddings, reducing computations to an acceptable level by controlling the embedding dimension $k$. We first establish the following theorem, which analyzes that the potential difference incurred by using our low-complexity modules in \textsf{CAM-AE} is negligibly small compared to the (computationally more expensive) original cross-attention \cite{vaswani2017attention}, defined as 
\begin{equation}
    Attention\left( {{\bf Q},{\bf K},{\bf V}} \right) := softmax \left( {\frac{{{\bf Q}{\bf K}^T }}{{\sqrt d }}} \right){\bf V} , 
\label{eq:origin_att}
\end{equation}
where ${\bf Q} = {\bf q} {\tilde {\bf W}}_\theta ^Q $, ${\bf K} = {\bf k}{\tilde{\bf W}}_\theta ^K  $, and ${\bf V} = {\bf v}{\tilde{\bf W}}_\theta ^V  $. Here, ${\bf q} \! \in \!\mathbb{R}^{\max \left\{ {\left| \mathcal{U} \right|,\left| \mathcal{I} \right|} \right\} \! \times \! d} $ and $\left\{ {{\bf k}, {\bf v}} \right\} \!\in \!\mathbb{R}^{\left| \mathcal{I} \right| \times  d} $ are the embedding matrices and $\left\{ {{\tilde{\bf W}}_\theta ^Q ,\!{\tilde{\bf W}}_\theta ^K ,\!\tilde{{\bf W}}_\theta ^V } \right\} \!\!\in \! \mathbb{R}^{^{d \times d} } $ are trainable parameters of $Attention $  in Eq. (\ref{eq:origin_att}).


\begin{corollary} 
\textit{Suppose that $\max \left\{ {\left| \mathcal{U} \right|,\left| \mathcal{I} \right|} \right\}$ is sufficiently large. If $k \ge {\rm{ }}{{5\!\ln \left( {\max \left\{ {\left| \mathcal{U} \right|,\left| \mathcal{I} \right|} \right\}} \right)} \mathord{\left/
 {\vphantom {{5\ln \left( {\left| I \right|} \right)} \!{\left( {\varepsilon ^2  - \varepsilon ^3 } \right)}}} \right.
 \kern-\nulldelimiterspace} {\left( {\varepsilon ^2  - \varepsilon ^3 } \right)}}{\rm{ }}
 $, then there exist matrices ${\bf E}_Q \in \mathbb{R}^{k \times \max \left\{ {\left| \mathcal{U} \right|,\left| \mathcal{I} \right|} \right\}} $, ${\bf E}_K ,{\bf E}_V \in \mathbb{R}^{k \times \left| \mathcal{I} \right|} $ and ${\bf D}  \in \mathbb{R}^{\left| \mathcal{I} \right| \times k} $ such that}
\begin{equation}
   \Pr \left( {\left| {\sqrt {\left\| {\frac{{{\bf D} \cdot softmax\left( {{\bf E}_Q {\bf A}{\bf E}_K^T } \right){\bf E}_V {\bf V}}}{{softmax\left( {\bf A} \right){\bf V}}}} \right\|}  - 1} \right| \le \varepsilon } \right) > 1 - o\left( 1 \right), 
\label{eq:theorem1}
\end{equation}
where ${\bf Q} \in \mathbb{R}^{\max \left\{ {\left| \mathcal{U} \right|,\left| \mathcal{I} \right|} \right\} \times d} $ and $\left\{ {{\bf K}, {\bf V}} \right\} \in \mathbb{R}^{\left| \mathcal{I} \right| \times d} $ are the embedding matrices in the original cross-attention; ${\bf A}=\frac{{{\bf Q}{\bf K}^T }}{{\sqrt d }}$; and $\varepsilon  > 0 $ is an arbitrarily small constant.
\end{corollary}

The proof of Theorem 1 is omitted due to page limitations. Theorem 1 implies that the probability that the two terms ${{softmax\left( {\bf A} \right){\bf V}}}$ and ${{\bf D} \! \cdot \! softmax\left( {{\bf E}_Q {\bf A}{\bf E}_K^T } \right){\bf E}_V {\bf V}}$ are approximately equal approaches one asymptotically when the maximum value of $|\mathcal{U}|$ and $|\mathcal{I}|$ is sufficiently large. We are capable of bridging this theorem and our \textsf{CAM-AE} model by setting ${\bf E}_V {\bf V} = {\bf V}_t$, ${\bf E}_K {\bf K} = {\bf K}_t $, ${\bf E}_Q {\bf Q} = {\bf Q}^{\left( h \right)} $, and ${\bf D}$ as in Eq. (\ref{eq:decoder}), where ${\bf E}_Q={\bf E}_h $, ${\bf E}_K={\bf E}_1 $, and ${\bf E}_V={\bf E}_1$, which leads to the conclusion that the term ${\bf D}\cdot{softmax\left( {{\bf E}_Q {\bf A}{\bf E}_K^T } \right)\!{\bf E}_V \!{\bf V}}$ in Eq. (\ref{eq:theorem1}) is equivalent to ${\bf D} \cdot  Attention_h\left( {{\bf Q}^{\left( h \right)},{\bf K}_t,{\bf V}_t} \right)$. From Theorem 1, one can see that the original cross-attention can be effectively approximated by our low-complexity modules in \textsf{CAM-AE}, which combines the cross-attention mechanism with {\it linear} transformations, thus significantly reducing the computational complexity (which is to be empirically validated later). In other words, we can control $k$ to maintain the amount of computations manageable while ensuring that the embeddings generated by our model closely approximate those from the original cross-attention, especially for large $\max \left\{ {\left| \mathcal{U} \right|,\left| \mathcal{I} \right|} \right\}$.

Additionally, to validate the scalability of the \textsf{CAM-AE} model, we analytically show its computational complexity during training by establishing the following theorem.
\begin{corollary}
    \textit{The computational complexity of \textsf{CF-Diff} training, including both the computation time of the forward-diffusion process and the training time of the reverse-denoising process, is given by $\mathcal{O}\left( {\max \left\{ {\left| \mathcal{U} \right|,\left| \mathcal{I} \right|} \right\}} \right) $.}
\end{corollary}
The proof of Theorem 2 is omitted due to page limitations. From Theorem 2, one can see that the computational complexity required to train \textsf{CF-Diff} scales {\it linearly} with the maximum between the number of users and the number of items. This is because we are capable of considerably reducing the computation of Eq. (\ref{eq:attention}) (corresponding to the cross-attention part in Figure \ref{fig:overview}) by controlling the embedding dimension $k$.

\section{Experimental Evaluation}

In this section, we systematically conduct extensive experiments to answer the following five key research questions (RQs):

\begin{itemize}
    \item {\bf RQ1}: How much does \textsf{CF-Diff} improve the top-$K$ recommendation over benchmark recommendation methods?
    \item {\bf RQ2}: How does each component in \textsf{CAM-AE} contribute to the recommendation accuracy?
    \item {\bf RQ3}: How many hops in \textsf{CF-Diff} benefit for the recommendation accuracy?
    \item {\bf RQ4}: How do key parameters of \textsf{CAM-AE} affect the performance of \textsf{CF-Diff}?
    \item {\bf RQ5}: How scalable is \textsf{CF-Diff} when the size of datasets increases?
\end{itemize}

\subsection{Experimental Settings}

{\bf Datasets.} We conduct our experiments on three real-world datasets widely adopted for evaluating the performance of recommender systems, which include MovieLens-1M (ML-1M)\footnote{https://grouplens.org/datasets/movielens/1m/.}, and two larger datasets, Yelp\footnote{https://www.yelp.com/dataset/.} and Anime\footnote{https://www.kaggle.com/datasets/CooperUnion/anime-recommendations-database.}. Table \ref{table:datasets} provides a summary of the statistics for each dataset.

\noindent\textbf{Competitors} To comprehensively demonstrate the superiority of \textsf{CF-Diff}, we present nine recommendation methods, including five general benchmark CF methods (NGCF \cite{wang2019neural}, LightGCN \cite{he2020lightgcn}, SGL \cite{wu2021self}, NCL \cite{lin2022improving}, and BSPM \cite{choi2023blurring}) and four generative model-based recommendation methods (CFGAN \cite{chae2018cfgan}, MultiDAE \cite{wu2016collaborative}, RecVAE \cite{shenbin2020recvae}, and DiffRec \cite{wang2023diffusion}).

\noindent\textbf{Performance metrics.} We follow the full-ranking protocol \cite{he2020lightgcn} by ranking all the non-interacted items for each user. In our study, we adopt two widely used ranking metrics, Recall@$K$ (R@$K$) and NDCG@$K$ (N@$K$), where $K \in \left\{ {10,20} \right\} $.

\noindent\textbf{Implementation details.} We use the best hyperparameters of competitors and \textsf{CF-Diff} obtained by extensive hyperparameter tuning on the validation set. We use the Adam optimizer \cite{kingma2014adam}, where the batch size is selected in the range of $\left\{ {32,64,128,256} \right\} $. In \textsf{CF-Diff}, the hyperparameters used in the diffusion model ({\it e.g.}, the noise scale $\beta_t$ and the diffusion step $T$) essentially follow the settings in \cite{wang2023diffusion}. We choose the best hyperparameters in the following ranges: $\left\{ {1,2,3,4} \right\} $ for the number of hops, $H$; $\left\{ {512,1024,2048} \right\} $ for the latent dimension $k$ in the attention-aided AE module; and $\left\{ {16,32,64,128} \right\} $ for the expanded dimension $d$, $\left\{ {1,2,3,4} \right\} $ for the number of layers, $N$, and $\left\{ {0.3,0.5,0.7} \right\} $ for $\alpha_h$'s in the multi-hop cross-attention module. All experiments are carried out with Intel (R) 12-Core (TM) E5-1650 v4 CPUs @ 3.60 GHz and GPU of NVIDIA GeForce RTX 3080. The code of \textsf{CF-Diff} is available at https://github.com/jackfrost168/CF\_Diff.





\begin{table}[t]
\small
  \captionsetup{skip=2pt}
  \caption{The statistics of three datasets.}
  \begin{tabular}{cccccl}
    \toprule
    Dataset & \# of users & \# of items & \# of interactions \\
    \midrule
    MovieLens-1M & 5,949 & 2,810 & 571,531 \\
    Yelp& 54,574 & 34,395 & 1,402,736  \\
    Anime & 73,515 & 11,200 & 7,813,737 \\
  \bottomrule
\end{tabular}
\label{table:datasets}
\vspace{-1em}
\end{table}

\begin{table*}[!t]\centering
\setlength\tabcolsep{5.0pt}
\small
  \captionsetup{skip=2.0pt}
  \caption{Performance comparison among \textsf{CF-Diff} and nine recommendation competitors for the three benchmark datasets. Here, the best and second-best performers are highlighted by bold and underline, respectively.}
  \label{tab:comparison}
  \begin{tabular}{c|cccc|cccc|cccc}
    \toprule[1pt]
    \multicolumn{1}{c|}{}&\multicolumn{4}{|c|}{ML-1M}&\multicolumn{4}{c|}{Yelp}&\multicolumn{4}{c}{Anime}\\
    \cmidrule{1-13}
           Method & R@10& R@20& N@10& N@20& R@10& R@20& N@10& N@20& R@10& R@20& N@10& N@20\\
    \midrule[1pt]
    NGCF & 0.0864& 0.1484& 0.0805&0.1008& 0.0428& 0.0726& 0.0255& 0.0345 & 0.1924& 0.2888 & 0.3515 & 0.3485\\
    LightGCN& 0.0824& 0.1419& 0.0793& 0.0982& 0.0505& 0.0858& 0.0312& 0.0417 & 0.2071 & 0.3043 & 0.3937 & 0.3824\\
    SGL& 0.0806& 0.1355& 0.0799& 0.0968& 0.0564& \underline{0.0944}& 0.0346& 0.0462 & 0.1994 & 0.2918 & 0.3748 & 0.3652\\
    NCL& 0.0878& 0.1471& 0.0819& 0.1011& 0.0535& 0.0906& 0.0326& 0.0438 & 0.2063 & 0.3047 & 0.3915 & 0.3819\\
    BSPM& 0.0884& 0.1494& 0.0750& 0.0957& \underline{0.0565}& 0.0932& 0.0331& 0.0439 & 0.2054 & \underline{0.3103} & 0.4355 & 0.4231\\
    \cmidrule{1-13}
    CFGAN& 0.0684& 0.1181& 0.0663& 0.0828& 0.0206& 0.0347 & 0.0129 &0.0172 & 0.1946 & 0.2889& 0.4601& 0.4289\\    
    MultiDAE& 0.0769& 0.1335& 0.0737& 0.0919& 0.0531& 0.0876& 0.0316& 0.0421 & 0.2142& 0.3085 & 0.4177 & 0.4125\\
    RecVAE& 0.0835& 0.1422& 0.0769& 0.0963& 0.0493& 0.0824& 0.0303& 0.0403 & 0.2137 & 0.3068 & 0.4105 & 0.4068\\
    DiffRec& \underline{0.1021}& \underline{0.1763}& \underline{0.0877}& \underline{0.1131}& 0.0554& 0.0914& \underline{0.0343}& \underline{0.0452} & \underline{0.2104} & 0.3012 & \underline{0.5047} & \underline{0.4649}\\
    \cmidrule{1-13}
    \textsf{CF-Diff}& {\bf 0.1077}& {\bf 0.1843}& {\bf 0.0912}& {\bf 0.1176}& {\bf 0.0585}& {\bf 0.0962}& {\bf 0.0368}& {\bf 0.0480} &{\bf 0.2191} & {\bf 0.3155} & {\bf 0.5152} & {\bf 0.4748}\\
    \bottomrule[1pt]
  \end{tabular}
  \vspace{-1.0em}
\end{table*}

\subsection{Results and Analyses}

In RQ1--RQ3, we provide experimental results on all datasets. For RQ4, we show here only the results on ML-1M in terms of N@$K$ due to space limitations, since the results on other datasets and metrics showed similar tendencies to those on ML-1M. Additionally, we highlight the best and second-best performers in each case of the following tables in bold and underline, respectively.

\subsubsection{Comparison with nine recommendation competitors ({\bf RQ1})}
We validate the superiority of \textsf{CF-Diff} over nine recommendation competitors through extensive experiments on the three datasets. Table \ref{tab:comparison} summarizes the results, and we make the following insightful observations.
\begin{enumerate}
    \item Our \textsf{CF-Diff} {\it consistently} and {\it significantly} outperforms all recommendation competitors regardless of the datasets and the performance metrics.
    \item The second-best performer tends to be DiffRec. Its superior performance among other generative model-based methods can be attributed to the use of diffusion models, known for their state-of-the-art performance in various fields. This enables DiffRec to more intricately recover user--item interactions for recommendations compared to VAE-based CF methods. However, DiffRec is consistently inferior to \textsf{CF-Diff}, primarily because it overlooks the high-order connectivity information, which is essential for capturing crucial collaborative signals.  
    \item The performance gap between \textsf{CF-Diff} ($X$) and Diffrec ($Y$) is the largest when the Yelp dataset is used; the maximum improvement rate of 7.29\% is achieved in terms of N@10, where the improvement rate (\%) is given by $\frac{{X - Y}}{Y} \times 100 $.
    \item Compared with GNN-based methods (NGCF, LightGCN, SGL, and NCL) that exploit high-order connectivity information through the message passing mechanism, our \textsf{CF-Diff} method exhibits remarkable gains. This superiority basically stems from the ability of inherently powerful diffusion models and avoiding the over-smoothing issue when integrating high-order connectivities.
    \item CFGAN shows relatively lower accuracies compared to other generative model-based methods. This performance degradation is caused by mode collapse during GAN training, resulting in inferior recommendation outcomes.
    
\end{enumerate}

\subsubsection{Impact of components in \textsf{CAM-AE} ({\bf RQ2})}

To discover what role each component plays in the success of our learning model, \textsf{CAM-AE}, we conduct an ablation study by removing or replacing each component in \textsf{CAM-AE}.

\begin{itemize}
    \item \textsf{CAM-AE}: corresponds to the original \textsf{CAM-AE} model.

    \item \textsf{CAM-AE-att}: removes the multi-hop cross-attention module in \textsf{CAM-AE}.

    \item \textsf{CAM-AE-ae}: removes the attention-aided AE in \textsf{CAM-AE}.

    \item \textsf{CAM-AE-self}: replaces the multi-hop cross-attention module in \textsf{CAM-AE} with the multi-hop {\it self-attention} module, which ignores the high-order connectivity information (by replacing ${\bf q}^{\left( h \right)}$ with ${\bf v}_t$).
\end{itemize}

The performance comparison among the original \textsf{CAM-AE} and its three variants is presented in Table \ref{tab:ablation} with respect to R@$K$ and N@$K$ on the three datasets. Our findings are as follows:
\begin{table}[t]\centering
\setlength\tabcolsep{5.0pt}
\small
  \captionsetup{skip=2pt}
  \caption{Performance comparison among \textsf{CAM-AE} and its three variants. Here, the best and second-best performers are highlighted by bold and underline, respectively.}
  \label{tab:ablation}
  \begin{tabular}{cc|cc|cc}
    \toprule[1pt]
    \cmidrule{1-6}
           Dataset& Method & R@10& R@20 & N@10 & N@20 \\
    \midrule[1pt]
    \multirow{4}*{\rotatebox{90}{ML-1M}} 
    & \textsf{CAM-AE-att} &0.1016& 0.1751& 0.0873& 0.1123\\
    & \textsf{CAM-AE-ae}& 0.1024& 0.1732& 0.0871& 0.1117\\
    & \textsf{CAM-AE-self}& \underline{0.1057}& \underline{0.1794}& \underline{0.0891}& \underline{0.1144}\\
    & \textsf{CAM-AE} & {\bf 0.1077}& {\bf 0.1843}& {\bf 0.0912}& {\bf 0.1176}\\
    \bottomrule[1pt]
    
    \multirow{4}*{\rotatebox{90}{Yelp}} 
    & \textsf{CAM-AE-att} &0.0553& 0.0905& 0.0342& 0.0448\\
    & \textsf{CAM-AE-ae}& OOM & OOM & OOM & OOM \\
    & \textsf{CAM-AE-self}& \underline{0.0574}& \underline{0.0952}& \underline{0.0355}& \underline{0.0469}\\
    & \textsf{CAM-AE} & {\bf 0.0585}& {\bf 0.0962}& {\bf 0.0368}& {\bf 0.0480}\\
    \midrule[1pt]

    \multirow{4}*{\rotatebox{90}{Anime}} 
    & \textsf{CAM-AE-att} &0.2091& 0.3024& 0.5023& 0.4623\\
    & \textsf{CAM-AE-ae}& OOM & OOM & OOM & OOM \\
    & \textsf{CAM-AE-self}& \underline{0.2112}& \underline{0.3094}& \underline{0.5079}& \underline{0.4678}\\
    & \textsf{CAM-AE} & {\bf 0.2191}& {\bf 0.3155}& {\bf 0.5152}& {\bf 0.4748}\\
    \bottomrule[1pt]
  \end{tabular}
  \vspace{-1.0em}
\end{table}
\begin{enumerate}
    \item The original \textsf{CAM-AE} always exhibits substantial gains over the other variants, which demonstrates that each component in \textsf{CAM-AE} plays a crucial role in enhancing the recommendation accuracy.
    \item \textsf{CAM-AE} outperforms \textsf{CAM-AE-att}, which can be attributed to the fact that the multi-hop cross-attention module is capable of infusing high-order connectivities into the proposed method to improve the performance of recommendations via CF.
    \item The performance gain of \textsf{CAM-AE} over \textsf{CAM-AE-ae} is relatively higher than that of the other variants for the ML-1M dataset. Additionally, the attention-aided AE's removal leads to out-of-memory (OOM) issues on the Yelp and Anime datasets, signifying its crucial role not only in extracting representations that can precisely capture the underlying patterns of user--item interactions but also in maintaining the computational complexity at acceptable levels.  
    \item \textsf{CAM-AE} is superior to \textsf{CAM-AE-self}. This confirms that infusing high-order connectivity information enriches the collaborative signal and thus results in performance enhancement even under a diffusion-model framework.
\end{enumerate}

\subsubsection{The impact of multi-hop neighbors ({\bf RQ3})}

To investigate how many hop neighbors in the \textsf{CF-Diff} method are informative, we present a variant of \textsf{CF-Diff}, \textsf{CF-Diff-$H$}, which considers up to $H$-hop neighbors {\it constantly} instead of optimally searching for the value of $H$ depending on a given dataset. The results are shown in Table \ref{tab:multi_hops} and our observations are as follows:
\begin{enumerate}
    \item \textsf{CF-Diff-3} outperforms \textsf{CF-Diff-2} on ML-1M and Yelp, indicating that incorporating a wider range of neighboring nodes into the \textsf{CAM-AE} model can positively influence the recommendation results through CF. 
    \item \textsf{CF-Diff-2} shows the highest recommendation accuracy on Anime. This means that 2-hop neighbors sufficiently capture the collaborative signal, and there is no need for exploiting higher-order connectivity information in this dataset.
    \item Notably, there is a decline in the performance of \textsf{CF-Diff-4}, because infusing 4-hop neighbors introduces an excess of global connectivity information. This surplus information potentially acts as noise, thereby interfering the personalized recommendations.
\end{enumerate}

\begin{table}[t]\centering
\setlength\tabcolsep{5.0pt}
\small
  \captionsetup{skip=2pt}
  \caption{Performance comparison according to different values of {\em H}. Here, the best and second-best performers are highlighted by bold and underline, respectively.}
  \label{tab:multi_hops}
  \begin{tabular}{cc|cc|cc}
    \toprule[1pt]
    \cmidrule{1-6}
           Dataset& Method & R@10& R@20 & N@10 & N@20 \\
    \midrule[1pt]
    \multirow{3}*{\rotatebox{90}{ML-1M}} 
    & \textsf{CF-Diff-2} &\underline{0.1062}& \underline{0.1786}& \underline{0.0907}& \underline{0.1164}\\
    & \textsf{CF-Diff-3} & {\bf 0.1077}& {\bf 0.1843}& {\bf 0.0912}& {\bf 0.1176}\\
    & \textsf{CF-Diff-4}& 0.1055& 0.1764& 0.0883& 0.1134\\
    \bottomrule[1pt]
    \multirow{3}*{\rotatebox{90}{Yelp}} 
    & \textsf{CF-Diff-2} &\underline{0.0572}& \underline{0.0935}& \underline{0.0351}& \underline{0.0462}\\
    & \textsf{CF-Diff-3} & {\bf 0.0585}& {\bf 0.0962}& {\bf 0.0368}& {\bf 0.0480}\\
    & \textsf{CF-Diff-4} & 0.0561 & 0.0917 & 0.0347 & 0.0455 \\
    \midrule[1pt]
    \multirow{3}*{\rotatebox{90}{Anime}} 
    & \textsf{CF-Diff-2} & {\bf 0.2191}& {\bf 0.3155}& {\bf 0.5152}& {\bf 0.4748}\\
    & \textsf{CF-Diff-3} & \underline{0.2082}& \underline{0.3021}& \underline{0.4998}& \underline{0.4586}\\
    & \textsf{CF-Diff-4} & 0.1938& 0.2824 & 0.4605 & 0.4236 \\
    \bottomrule[1pt]
  \end{tabular}
  \vspace{-2.3em}
\end{table}

\subsubsection{The effect of hyperparameters ({\bf RQ4})}
We analyze the impact of key parameters of \textsf{CAM-AE}, including $k$, $d$, $N$, and $\alpha_h$, on the recommendation accuracy for the ML-1M dataset. In this experiment, we consider 3-hop neighbors ({\it i.e.}, $H=3$). For notational convenience, we denote $\alpha_2=\alpha$ and $\alpha_3=1-\alpha$, which signify the importance of 2-hop and 3-hop neighbors, respectively. When a hyperparameter varies so that its effect is clearly revealed, other parameters are set to the following pivot values: $k = 500,d = 16,N = 2,\alpha  = 0.7 $. Our findings are as follows:

({\bf Effect of $k$}) From Figure \ref{fig:k_d}a, the maximum N@10 and N@20 are achieved at $k = 500$ on ML-1M. It reveals that high values of $k$ degrade the performance since the resulting embeddings contain more noise and low values of $k$ result in insufficient information during training. Hence, it is crucial to suitably determine the value of $k$ in guaranteeing satisfactory performance. 

({\bf Effect of $d$}) From Figure \ref{fig:k_d}b, the maximum N@10 and N@20 are achieved at $d = 16$ on ML-1M. Using values of $d$ that are too high and too low has a negative impact on the model's expressiveness. Thus, it is important to appropriately determine the value of $d$ depending on the datasets.

({\bf Effect of $N$}) From Figure \ref{fig:n_alpha}a, the maximum N@10 and N@20 are achieved at $N = 2$ on ML-1M. A higher $N$ rather degrades the performance, possibly due to the over-fitting problem. Thus, the value of $N$ should be carefully chosen based on given datasets.

({\bf Effect of $\alpha$}) Figure \ref{fig:n_alpha}b shows that the maximum N@10 and N@20 are achieved at $\alpha = 0.7$ on ML-1M. Tuning $\alpha$ is crucial since it directly determines the model's ability while balancing between neighbors that are different hops away from the target user, which in turn affects the recommendation performance.

\begin{figure}[t!]
\captionsetup{skip=3pt}
\pgfplotsset{footnotesize,samples=10}
\centering

\begin{tikzpicture}[scale=0.97]
\begin{axis}[
xlabel= (a) Effect of $k$, ylabel = NDCG,
xlabel style={yshift=0.8em}, 
ylabel style={yshift=-1.8em}, 
width = 0.24\textwidth, height = 3.7cm,
xmin=300,xmax=1000,ymin=0.08,ymax=0.13,
xtick={350,500,650,800,950},
ytick={0.08,0.10,0.12},
yticklabel style={/pgf/number format/fixed},
]
    \addplot+[color=fig_7_color_3] coordinates{(350, 0.0863)(500, 0.0915)(650, 0.0912)(800, 0.0907)(950, 0.0862)};
    \addplot+[color=fig_7_color_4] coordinates{(350, 0.1123)(500, 0.118)( 650, 0.1176)(800, 0.1164)(950, 0.112)};
\end{axis}
\end{tikzpicture}%
\begin{tikzpicture}[scale=0.97]
\begin{axis}[
xlabel= (b) Effect of $d$, ylabel = NDCG,
xlabel style={yshift=0.8em}, 
ylabel style={yshift=-1.8em}, 
width = 0.24\textwidth, height = 3.7cm,
xmin=6,xmax=36,ymin=0.08,ymax=0.13,
xtick={8,16,24,32},
ytick={0.08,0.10,0.12},
yticklabel style={/pgf/number format/fixed},
legend style={at={(1.05,0.7)},anchor=north west,legend columns=1, font=\footnotesize}, 
]
    \addplot+[color=fig_7_color_3] coordinates{(8, 0.0908)(16, 0.0912)(24, 0.0872)(32, 0.087)};
    \addplot+[color=fig_7_color_4] coordinates{(8, 0.1167)(16, 0.1176)(24, 0.1127)(32, 0.1125)};
    \legend{N@10,N@20}
\end{axis}
\end{tikzpicture}

\captionsetup{skip=1.0pt}
\caption{The effect of hyperparameters $k$ and $d$ on N@K for the ML-1M dataset.}
\label{fig:k_d}
\vspace{-1em}
\end{figure}

\begin{figure}[t!]
\captionsetup{skip=0pt}
\pgfplotsset{footnotesize,samples=10}
\centering
\begin{tikzpicture}[scale=0.97]
\begin{axis}[
xlabel= (a) Effect of $N$, ylabel=NDCG,
xlabel style={yshift=0.8em}, 
ylabel style={yshift=-1.7em}, 
width=0.24\textwidth, height=3.7cm,
xmin=0, xmax=5, ymin=0.05, ymax=0.13,
xtick={1,2,3,4},
ytick={0.06, 0.08,0.10, 0.12},
ybar,
bar width=7pt,
xticklabel style={yshift=3pt},
yticklabel style={/pgf/number format/fixed},
]
\draw[dashed] (axis cs:0.0,0.0915) -- (axis cs:2,0.0915);
\addplot+[ybar, color=fig_7_color_1,bar shift=-3.5pt,draw=black, line width=1.2pt] coordinates{(1, 0.0868)(2, 0.0912)(3, 0.0883)(4, 0.0587)};
\draw[dashed] (axis cs:0.0,0.1180) -- (axis cs:2,0.1180);
\addplot+[ybar, color=fig_7_color_2, bar shift=2.5pt, draw=black, line width=1.2pt] coordinates{(1, 0.1119)(2, 0.1176)(3, 0.1131)(4, 0.0726)};
\end{axis}
\end{tikzpicture}
\begin{tikzpicture}[scale=0.97]
\begin{axis}[
xlabel= (b) Effect of $\alpha$, ylabel=NDCG,
xlabel style={yshift=0.8em}, 
ylabel style={yshift=-1.7em}, 
width=0.24\textwidth, height=3.7cm,
xmin=0.15, xmax=1, ymin=0.05, ymax=0.13,
xtick={0.3, 0.5, 0.7,0.9},
ytick={0.06, 0.08, 0.1, 0.12},
ybar,
bar width=7pt,
xticklabel style={yshift=3pt},
yticklabel style={/pgf/number format/fixed},
legend style={at={(1.05,0.7)},anchor=north west,legend columns=1,font=\footnotesize}, 
]
    \draw[dashed] (axis cs:0.0,0.0925) -- (axis cs:0.7,0.0925);
    \addplot+[ybar, color=fig_7_color_1, bar shift=-3.5pt,draw=black, line width=1.2pt] coordinates{(0.3, 0.0890)(0.5, 0.0902)(0.7, 0.0921)(0.9, 0.0902)};
    \draw[dashed] (axis cs:0.0,0.1188) -- (axis cs:0.7,0.1188);
    \addplot+[ybar, color=fig_7_color_2, bar shift=2.5pt, draw=black, line width=1.2pt] coordinates{(0.3, 0.1142)(0.5, 0.1156)(0.7, 0.1180)(0.9, 0.1153)};
    \legend{N@10,N@20}
\end{axis}
\end{tikzpicture}

\captionsetup{skip=1pt}
\caption{The effect of hyperparameters $N$ and $\alpha$ on N@K for the ML-1M dataset.}
\label{fig:n_alpha}
\end{figure}


\begin{figure}[t!]
\vspace{-1em}
\captionsetup{skip=0pt}
\pgfplotsset{footnotesize,samples=10}
\centering
\begin{tikzpicture}[scale=0.95]
    \begin{axis}[        
    xlabel= (a) $\left| \mathcal{U} \right|$, 
    ylabel= Execution time (s), 
    xlabel style={yshift=1.1em}, 
    ylabel style={yshift=-2.1em}, 
    grid=major, grid style={dashed}, width = 4.9cm, height = 4.0cm , legend style={at={(0.41,0.2)},anchor=west},
    xtick={10000,35000,65000,90000},
    scaled x ticks=false,
    xticklabel style={font=\scriptsize},
    xticklabels={$1 \cdot 10^4$,$3 \cdot 10^4$,$6 \cdot 10^4$,$9 \cdot 10^4$},
    xticklabel style={yshift=3pt},
    yticklabel style={font=\scriptsize},
    ],
        \addplot [color=fig_8_color_1, mark=square, legend = \textsf{CF-Diff}]
            coordinates {
                (10000, 10.25)(30000, 30.5)(40000,41.25)(60000, 63.75)(70000, 76)(80000, 89)(90000, 101.25)
            };
        \addplot [ dashed, color=red,        mark=none, legend = $$\mathcal{O}\left( {\left| \mathcal{U} \right|} \right)$$ ]
            coordinates {
                (5000,4.25)(90000,101.25)
            };
    \addlegendentry{\textsf{CF-Diff}}
    \addlegendentry{$\mathcal{O}\left( {\left| \mathcal{U} \right|} \right)$}
    \end{axis}
\end{tikzpicture}
\begin{tikzpicture}[scale=0.95]
    \begin{axis}[        
    xlabel= (b) $\left| \mathcal{I} \right|$, 
    ylabel= Execution time (s), 
    xlabel style={yshift=1.1em}, 
    ylabel style={yshift=-2.1em}, 
    grid=major, grid style={dashed}, width = 4.9cm, height = 4.0cm, legend style={at={(0.43,0.2)},anchor=west},
    scaled x ticks=false,
    xtick={10000,70000,120000,188000},
    xticklabel style={font=\scriptsize},
    xticklabels={$1 \cdot 10^4$,$6 \cdot 10^4$,$13 \cdot 10^4$,$20 \cdot 10^4$},
    xticklabel style={yshift=3pt},
    yticklabel style={font=\scriptsize},
    ]
        \addplot [color=fig_8_color_1, mark=square, legend = \textsf{CF-Diff}]
            coordinates {
                (10000, 10.5)(40000, 12.25)(60000,13.1315)(80000, 14.25)(120000, 16.25)(160000, 18.25)(200000, 20.5)
            };
        \addplot [ dashed, color=red,        mark=none, legend = $$\mathcal{O}\left( {\left| \mathcal{I} \right|} \right)$$ ]
            coordinates {
                (5000,10.2368)(200000,20.4999)
            };
    \addlegendentry{\textsf{CF-Diff}}
    \addlegendentry{$\mathcal{O}\left( {\left| \mathcal{I} \right|} \right)$}
    \end{axis}
\end{tikzpicture}
\captionsetup{skip=1pt}
\caption{The computational complexity of \textsf{CF-Diff}, where the plots of the execution time versus $\left| \mathcal{U} \right|$ in Figure 6a and the execution time versus $\left| \mathcal{I} \right|$ in Figure 6b are shown.}
\label{fig:time}
\vspace{-1.0em}
\end{figure}

\subsubsection{Computational complexity ({\bf RQ5})}
To empirically validate the scalability of our \textsf{CF-Diff} method, we measure the execution time during training on synthetic datasets having user--item interactions. These interactions are generated purely at random, simulating a sparsity level of 0.99, analogous to that observed on Yelp and Anime. By setting different $\left| \mathcal{U} \right|$'s and $\left| \mathcal{I} \right|$'s, we can create user--item interactions of various sizes. More specifically, we generate two sets of user--item interactions: in the first set, we generate a set of interactions with $\left| \mathcal{I} \right| = 1e^4$ and $\left| \mathcal{U} \right| = \left\{ {1e^4 ,3e^4 ,4e^4 ,6e^4 ,7e^4 ,8e^4 ,9e^4 } \right\} $; and in the second set, we generate another set of interactions with $\left| \mathcal{U} \right| = 1e^4$ and $\left| \mathcal{I} \right| = \left\{ {1e^4 ,4e^4 ,6e^4 ,8e^4 ,12e^4 ,16e^4, 20e^4 } \right\} $. Figure \ref{fig:time}a ({\it resp.} Figure \ref{fig:time}b) illustrates the execution time (in seconds) per iteration of \textsf{CF-Diff}, including the forward-diffusion process and the reverse-denoising process, as the number of users ({\it resp.} the number of items) increases. The dashed line indicates a linear scaling in $\left| \mathcal{U} \right|$ and $\left| \mathcal{I} \right|$, derived from Theorem 2. It can be seen that our empirical evaluation concurs with the theoretical analysis.

\section{Related Work}

In this section, we review some representative methods in two broad fields of research, including 1) benchmark CF methods and 2) generative model-based recommendation methods.

\subsection{General Benchmark CF}
The most common paradigm of CF is to factorize the user--item interaction matrix into lower-dimensional matrices \cite{mnih2007probabilistic, koren2009matrix, rendle2012bpr}. The dot product in MF can be replaced with a multi-layer perceptron (MLP) to capture the non-linearities in the complex behavior of such interactions \cite{he2017neural, chen2020efficient}. To analyze beyond direct user connections to items, high-order connectivities are essential for understanding the user preferences, leading to the rise of GNNs in CF for modeling these complex relationships \cite{berg2017graph}. GC-MC \cite{berg2017graph} first proposed a graph AE framework for recommendations using message passing on the user--item bipartite graph. NGCF \cite{wang2019neural} employed GNNs to propagate user and item embeddings on the bipartite graph capturing the collaborative signal in complex high-order connectivities. NIA-GCN \cite{sun2020neighbor} was developed by taking into account both the relational information between neighboring nodes and the heterogeneous nature of the user--item bipartite graph. LightGCN \cite{he2020lightgcn} improved the performance by lightweight message passing, omitting feature transformation and nonlinear activation. UltraGCN \cite{mao2021ultragcn} advanced efficiency by skipping infinite layers of explicit message passing and directly approximating graph convolution limits with a constraint loss. BSPM \cite{choi2023blurring} made a connection between the concept of blurring-sharpening process models and graph filtering \cite{shen2021powerful}, utilizing ordinary differential equations to model the perturbation and recovery of user--item interactions. Additionally, contrastive learning was used to further improve the recommendation accuracy by taking node self-discrimination into account \cite{wu2021self, lin2022improving}.

\subsection{Generative Model-Based Recommendation}

{\bf GAN-based methods.} Generative adversarial network (GAN)-based models in CF employ a generator to estimate user--item interaction probabilities, optimized through adversarial training \cite{goodfellow2014generative, wang2017irgan, guo2020ipgan, wu2019pd}. RecGAN \cite{bharadhwaj2018recgan} combined recurrent neural network (RNN) with GAN for capturing complex user--item interaction patterns, while CFGAN \cite{chae2018cfgan} enhanced the recommendation accuracy with real-valued vector-wise adversarial learning. Nevertheless, adversarial training is often associated with training instability and mode collapse, potentially leading to suboptimal performance \cite{arjovsky2017towards, metz2016unrolled}.

{\bf VAE-based methods.}
The denoising AE (DAE) was firstly used for top-$K$ recommendations, learning latent representations from corrupted user preferences \cite{wu2016collaborative}. CVAE \cite{li2017collaborative} extended this by using a VAE to learn latent representations of items from ratings and multimedia content for multimedia recommendations. A series of VAE-based methods \cite{liang2018variational, ma2019learning, shenbin2020recvae} were further developed for CF with implicit feedback, enhancing the accuracy, interpretability, and robustness by incorporating a multinomial likelihood and a Bayesian approach for user preference modeling. However, VAE-based models struggle to balance between simplicity and representations of complex data, with simpler models possibly failing to capture diverse user preferences and more complex models potentially being computationally intractable \cite{sohl2015deep}.

{\bf Diffusion model-based methods.} Recently, diffusion models have achieved state-of-the-art performance in image generation by decomposing the image generation process into a series of DAEs. CODIGEM \cite{walker2022recommendation} extended this with the denoising diffusion probabilistic model (DDPM) in \cite{ho2020denoising} to recommender systems, leveraging the intricate and non-linear patterns in the user--item interaction matrix. Additionally, diffusion models have been successfully applied to sequential recommendations \cite{yang2023generate, liu2023diffusion, wu2023diff4rec,li2023diffurec}. Inspired by score-based generative models \cite{song2020score},  DiffRec \cite{wang2023diffusion} accommodated diffusion models to predict unknown user--item interactions in a denoising manner by gradually corrupting interaction histories with scheduled Gaussian noise and then recovering the original interactions iteratively through a neural network.

{\bf Discussion.} Despite the impressive performance of current diffusion model-based recommender systems, existing models overlook high-order user--item connectivities that reveal co-preference patterns between users and items. These high-order connectivities among users and items are crucial in CF performed with limited direct user--item interactions, aiding in delivering more precise and personalized recommendations. However, effectively incorporating such high-order connectivity information remains a significant challenge in diffusion model-based CF.

\section{Conclusions}

In this paper, we explored an open yet fundamental problem of how to empower CF-based recommender systems when diffusion models are employed as a core framework for training. To tackle this challenge, we proposed \textsf{CF-Diff}, a diffusion model-based approach for generative recommender systems, designed to infuse high-order connectivity information into our own learning model, \textsf{CAM-AE}, while preserving the model's complexity at manageable levels. Through extensive experiments on three real-world benchmark datasets, we demonstrated (a) the superiority of \textsf{CF-Diff} over nine state-of-the-art recommendation methods while showing dramatic gains up to 7.29\% in terms of NDCG@10 compared to the best competitor, (b) the theoretical findings that analytically confirm the computational tractability and scalability of \textsf{CF-Diff}, (c) the effectiveness of core components in \textsf{CAM-AE}, and (d) the impact of tuning key hyperparameters in \textsf{CAM-AE}.

\begin{acks}
This work was supported by the National Re- search Foundation of Korea (NRF), Republic of Korea Grant by the Korean Government through MSIT under Grants 2021R1A2C3004345 and RS-2023-00220762 and by the Institute of Information and Communications Technology Planning and Evaluation (IITP), Republic of Korea Grant by the Korean Government through MSIT (6G Post-MAC--POsitioning and Spectrum-Aware intelligenT MAC for Computing and Communication Convergence) under Grant 2021-0-00347.
\end{acks}

\bibliographystyle{ACM-Reference-Format}
\balance
\bibliography{sample-base}

\end{document}